# Comparison of 5G Performance Post-Merger between Two Network Operators Using Field Tests in Urban Areas

Surachai Chatchalermpun[1], Therdpong Daengsi[2], Pakkasit Sriamorntrakul[2],
Kritphon Phanrattanachai[3]
[1]Department of Computer Engineering, Faculty of Engineering, South - East Asia University, Thailand
[2]Sustainable Industrial Management Engineering, Faculty of Engineering, Rajamangala University of Technology Phra Nakhon, Thailand
[3]Department of Computer Engineering, Faculty of Agricultural & Industrial Technology, Phetchabun Rajabhat University, Thailand

**ABSTRACT**

In late Q1/2023, DTAC and TRUE officially completed their merger. Consequently, this study was initiated to ascertain whether their respective 5G networks had been seamlessly integrated several months following the merger. The investigation involved conducting drive tests along two predefined routes within the urban areas of Bangkok, employing the G-NetTrack Pro tool for testing and data collection. Additionally, stationary tests were conducted in two crowded places using an application called Speedtest. Subsequently, an array of Quality of Service (QoS) metrics, including Reference Signal Received Power (RSRP), Reference Signal Received Quality (RSRQ), Signal to Noise Ratio (SNR), Download (DL), Upload (UL) speeds, and latency, were meticulously analyzed and presented. The findings of this study unveiled that, despite the successful completion of the DTAC and TRUE merger from a business standpoint, the technical integration of their respective 5G networks had not been finalized, although there were no significant differences between DTAC and TRUE for DL (p-value = 0.542) and UL (p-value = 0.090). Notably, significant differences were found between DTAC and TRUE for four metrics, including RSRP, RSRQ, SNR, and latency (p-values < 0.05). Remarkably, roaming functionalities were still operational between the two networks.

*Corresponding Author:*

Kritphon Phanrattanachai
Department of Computer Engineering, Faculty of Agricultural & Industrial Technology, Phetchabun Rajabhat University
83 M.11, Sadiang, Muang, Phetchabun, Thailand
Email: kritphon.ai@pcru.ac.th

## 1. INTRODUCTION

To survive in business competition and to react to changes in the competitive landscape, a merger may be used to lead to changes in the structure of the combined company and its performance [1]. A merger is defined in [2] as a combination of at least two companies into one, during which no new investments are made and the merging organization loses its identity. In the past decades, there have been mergers of businesses in ASEAN countries. For example, the merger of financial firms in Thailand and Indonesia [3]-[4].

In the ICT industry, a few years ago, due to intense competition and razor-thin profit margins among players, there was a merger between T-Mobile, the third-largest carrier, and Sprint, the fourth-largest one in the USA [5]. While in Thailand, TRUE and DTAC made the strategic decision to merge in Q4/2021 [6]. However, the merger process between these two prominent mobile network operators was far from straightforward and required approximately 1.5 years to complete, as depicted in Figure 1 [7]. After the merger, as mentioned in [8], the merged company, called True Corp for short, became the largest mobile network operator in Thailand. It occupies several frequency spectrum bands, including 700 MHz, 850 MHz, 900 MHz, 1800 MHz, 2100 MHz, 2300 MHz, and 2600 MHz, which covers a bandwidth of 1,350 MHz in total. However, after the business merger was completed, the technical merger was still questionable since some consumers, who are the customers of these old mobile network operators (MNOs)—TRUE and DTAC—were concerned about the quality of services provided by the merged company, particularly the old DTAC customers. Therefore, this study was conducted to investigate the status of the technical merger and whether it has been completed.

This paper is the extended version of [8]. However, the advancement beyond the previous version is adding the stationary tests. This article consists of four sections. The first section is the introduction, including background, objectives, 5G overview, and literature review. The second one is the methodology. In the third section, it covers the results and the analysis. Lastly, in Section 4, the discussion and conclusion are presented. This section especially covers the major contribution of this study associated with the disclosure of the status of the technical merger between two old operators.

According to the objectives of this study, although TRUE and DTAC were merged completely in terms of law and business, it is questionable whether the 5G mobile networks were also completely merged or not. Therefore, this study using drive tests was conducted to fulfill the curiosity of former TRUE and DTAC customers who may be confused about the uncertain situation after the business merger. The gathered data consists of six QoS parameters, consisting of download (DL) and upload (UL) speeds, Reference Signal Received Power (RSRP), Reference Signal Received Quality (RSRQ), Signal to Noise Ratio (SNR), and latency. Furthermore, this study can fulfill the research gaps associated with 5G field trial studies in Thailand, which were mostly conducted in stationary mode and did not cover the issues mentioned in this study.

In order to understand about this study clearly, the background about 5G technology must be described. The fifth-generation wireless technology for mobile telecommunications or 5G technology, began its operations in various countries in 2019. The key features of 5G are flexibility, scalability, adaptability, and realizable network development [9]. 5G networks support three use cases defined by ITU, consisting of Enhanced Mobile Broad-Band (EMBB), Massive Machine-Type Communications (MMTC), and Ultra Reliable Low Latency Communications (URLLC) [10]. This means that 5G networks can support a wider range of technologies, such as machine to machine communications, internet of things, and device to device communications [11]. Furthermore, 5G wireless technologies consume less power than previous technologies, since energy consumption is a critical concern for 5G MNOs [12], [13]. Thus, enhanced technologies, such as deep sleep and symbol aggregation shutdown, have been developed and utilized to reduce operational costs and carbon emissions [13], [14]. Moreover, one of the advancements beyond 4G or LTE is high reliability in privacy and security from network slicing [15], [16].

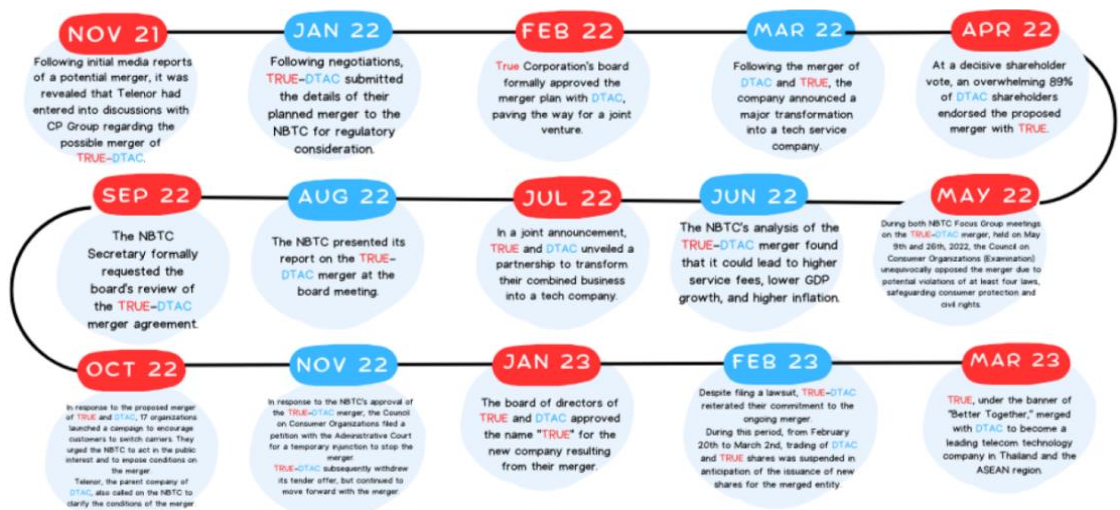

Figure 1. Timeline of the merger between DTAC and TRUE [7]

In the realm of 5G, all devices are characterized by their intelligence, connectivity, and responsiveness. These qualities enable distributed intelligence and collaborative computing, underpinned by the use of several technologies, including device-to-device, cognitive radio, software-defined networking, and massive multi-input, multi-output [17]. The capabilities of 5G extend to the establishment of high-density hetero networks, with the capacity to support up to one million devices per square kilometer. This groundbreaking technology promises a peak download rate of up to 20 Gbps in theory, which is significantly beyond 4G or LTE performance and can deliver a peak download of about 300 Mbps [18]. To support high data rate services, high frequencies, and wide bandwidths are required. Thus, 5G new radio (NR) bands in the sub-6 short for gigahertz (GHz) spectrum were allocated for 5G networks, However, the bands above 6 GHz are also allocated but that band is still encouraged to be designed, developed and implemented [19].

Based on ITU-T standard, 5G technology is interoperable with previous technologies, including 4G or LTE [20]. That means, when 5G network mode is not available, 5G user equipment or smartphones will connect 4G or LTE network instead automatically. Thus, 5G devices can be used for communications still. Furthermore, 5G supports two modes, stand-alone (SA) that uses 5G NR and 5G core networks and Non-standalone (NSA) that uses 5G NR with some parts of 4G or LTE core networks, for services [21].

In addition, to understand about this study, the background about QoS parameters must be also presented in detail. To evaluate network performance, numerous network parameters are usually considered and investigated. However, in this paper, six QoS parameters as follows were focused:

1) RSRP [22]: it measures the strength of the signal received from the Base Station (BS) equipment to the User Equipment (UE). Typically, a stronger signal is received when the UE is closer to the BS. RSRP values are typically categorized into five categories, consisting of excellent (-80 to -44 dBm), good (-90 to -80 dBm), fair (-100 to -90 dBm), poor (-110 to -100 dBm), and very poor (-140 to -110 dBm).

2) RSRQ [22]: it reflects the overall quality of the signal received by the UE, considering signal strength, noise, and interference. RSRQ values are typically categorized as excellent (-10 to -3 dBm), good (-12 to -10 dBm), fair (-14 to -12 dBm), poor (-17 to -14 dBm), and very poor (-20 to -17 dBm).

3) SNR [22]: it measures the ratio between the average received power and the average interference and noise. It plays a role in wireless communications, especially with advanced technologies. SNR values are categorized as excellent (10 to 30 dBm), good (3 to 10 dBm), fair (0 to 3 dBm), and poor (-20 to 0 dBm).

4) Data speeds (DL and UL) [23]: these are two critical parameters for network performance. They are easily understood by users, with higher values indicating better performance. Generally, download speeds tend to be faster than upload speeds.

5) Latency [24], [25]: often referred to as "delay," is a vital parameter that can significantly impact network performance. It is typically expressed as half of the "ping" values measured by various applications. Round-trip latency includes factors such as data scheduling latency, data transmission processing latency, data transmission latency, and data propagation latency. It is very important for real-time applications and services, including mission-critical applications (e.g., remote surgery), entertainment applications and online games.

These six QoS parameters provide essential insights into network performance and play a crucial role in ensuring a high-quality user experience, particularly in the context of emerging technologies like 5G. Monitoring and optimizing these parameters are essential for delivering the demanding quality of service (QoS) expected by users and applications.

Last but not least, there are many previous works associated with 5G technology and applications, nevertheless, only the related works based on 5G that used the G-NetTrack Pro applications and a few works associated with the QoS parameters and mobility tests were studied in detail and summarized in Table 1 [23], [26-36]. However, from the survey, there is no recent research work focusing on mobility tests using the G-NetTrack Pro application. Therefore, this can be the research gap that can be fulfilled and can be applied to investigation the status of the 5G system merger between DTAC and TRUE.

## 2. METHOD

This study was designed to use not only mobility tests but also stationary tests, which are field tests, for data gathering. Mobility tests is matched with one special application, while stationary test can be conducted with other application. Further details associated with both kinds of tests can be described Sub-sections 2.1 and 2.2, respectively.

❒ 5

Table 1. Related works

| Ref. | Technology | | | Mode | | QoS Parameter | | | | | | | | | Country | Tools | Major Findings / Major Contributions |
| | 5G | 4G/LTE | 3G | Stationary | Mobility | DL | UL | Latency/Ping | Loss | Jitter | RSRI | RSRP | RSRQ | SNR | | | |
| --- | --- | --- | --- | --- | --- | --- | --- | --- | --- | --- | --- | --- | --- | --- | --- | --- | --- |
| [23] | ✓ | - | - | ✓ | ✓ | ✓ | ✓ | - | - | - | - | - | - | - | Thailand | SpeedTest | They conducted field trials, and they found that the 5G quality of service in Bangkok measured in 2023 is worse than the quality measured in 2021. |
| [26] | ✓ | ✓ | - | ✓ | ✓ | ✓ | ✓ | - | - | - | ✓ | ✓ | ✓ | ✓ | Ireland | G-NetTrack Pro | They constructed a model for 5G NR physical layer round trip latency. The paper provides an analysis and estimation of the average physical layer latency. |
| [27] | ✓ | ✓ | ✓ | - | ✓ | ✓ | ✓ | ✓ | - | ✓ | ✓ | ✓ | ✓ | - | Oman | G-NetTrack Pro | They revealed that the average throughput in Ibra City, including the downlink speed of 20 Mbps and the uplink speed of 15 Mbps, is 36.5 ms, while the minimum average ping is 36.5 ms. |
| [28] | ✓ | ✓ | - | - | ✓ | - | - | - | ✓ | - | ✓ | ✓ | ✓ | ✓ | Indonesia | G-Net Track Pro | They discussed that the presence of buildings, tree heights, network density, and weather cause propagation and then signal strength. The long distance from the transmitter to the receiver is also a negative factor for the 4G LTE network. |
| [29] | ✓ | - | - | - | ✓ | ✓ | ✓ | ✓ | ✓ | ✓ | - | - | - | - | Indonesia | G-NetTrack Pro & SpeedTest | They revealed the current capabilities and performance of Telkomsel's 5G network in the measured area in Indonesia. Telkomsel's network shows low latency, no packet loss, and satisfactory data throughput, which are good for providing reliable and efficient 5G services. |
| [30] | ✓ | ✓ | - | ✓ | ✓ | ✓ | ✓ | - | ✓ | - | ✓ | ✓ | ✓ | | Indonesia | G-NetTrack Pro & SpeedTest | They revealed that the 5G network in Jakarta has better signal and service quality than the 4G/LTE network, while the 5G network has a coverage radius of approximately 1 km per base transceiver station. |
| [31] | ✓ | ✓ | - | - | ✓ | ✓ | ✓ | - | - | - | - | - | - | ✓ | Romania | StarTrinity CSTA | They unveiled the current state of the cellular network implemented in Iasi City for automotive applications. They proposed methods for data acquisition, processing, and visualization to highlight favorable or unfavourable characteristics of V2X communications. |
| [32] | ✓ | - | - | - | - | - | - | ✓ | - | - | - | - | - | - | Oman | G-NetTrack Pro | They presented that the paper analyses the performance of mobile broadband cellular networks in Cyberjaya City, Malaysia, including signal quality, throughput, ping, and handover. |
| [33] | ✓ | - | - | - | ✓ | - | - | ✓ | ✓ | - | - | - | - | - | Portugue & Spain | VRU and tcpdump | They found that the 5G network supports low latency for VRU safety in the real testbed, while GPS accuracy in commercial smartphones limits the usefulness of positioning data. |
| [34] | ✓ | ✓ | - | ✓ | - | ✓ | - | ✓ | - | - | - | ✓ | ✓ | - | China | Speedtest servers | They unveiled that 5G links can approach Gbps throughput, but legacy TCP limits capacity. Whereas latency is too high for tactile applications, power consumption escalates over 4G. |
| [35] | ✓ | ✓ | - | - | - | - | - | ✓ | - | - | - | - | - | - | South Korea | Speedtest | They presented that a 5G edge server enables TCP algorithms to achieve 1.8 Gbps throughput. They found that TCP receive buffer size limits 5G network performance, not 4G, while 5G edge benefits download-centric apps more than CPU-intensive or uplink-heavy tasks. |
| [36] | ✓ | - | - | ✓ | - | ✓ | ✓ | ✓ | - | - | - | - | - | - | France, Germany & Italy | Speedtest servers | They presented that roaming in the 5G era leads to decreased throughput and increased latency. Users experience higher throughput under 5G coverage while roaming. |

## 2.1 Mobility tests

Drive tests, which are sub-types of mobility tests were selected for this study because they can cover a wide area within a short period. Drive tests were conducted from August 30, 2023 (10:30 p.m.) to August 31, 2023 (1:00 a.m.) since the traffic was clear. There were two routes in this study, the first route called Route 1 (R1) is the inner ring, while the second route called Route 2 (R2), is the outer ring, as presented in Figure 2. The inner ring in this study is the expressway within the heart of Bangkok, while the outer ring is the roads that encompass the urban areas of Bangkok. When the drive tests were conducted, two smartphones with the application called G-NetTrack Pro (subscribed version—not free) were put on the front seat of the car beside the driver. Two 5G smartphones (see Figure 3), with the same model from the same vendor or manufacturer, used Android 13 as an operating system (OS) and had a CPU of Snapdragon 695 5G/octa-core and a memory of 128GB/8GB RAM. This study utilized an application called G-NetTrack Pro as mentioned in [26]-[30], [33] was installed on both smartphones.

After the data gathering using mobility tests and services provided by TRUE and DTAC, the data were processed as presented in Figure 4. Since this study is intended to be mobility tests, the data collected during the car stop were removed. Next, the data were filtered for DL data, UL data, signal strength data, and latency data. Then, the outliers were considered and then gotten rid of. Finally, the prepared data were presented in the next section. Furthermore, statistical analysis was also conducted to investigate whether there were significant differences between the results from the two MNOs or not.

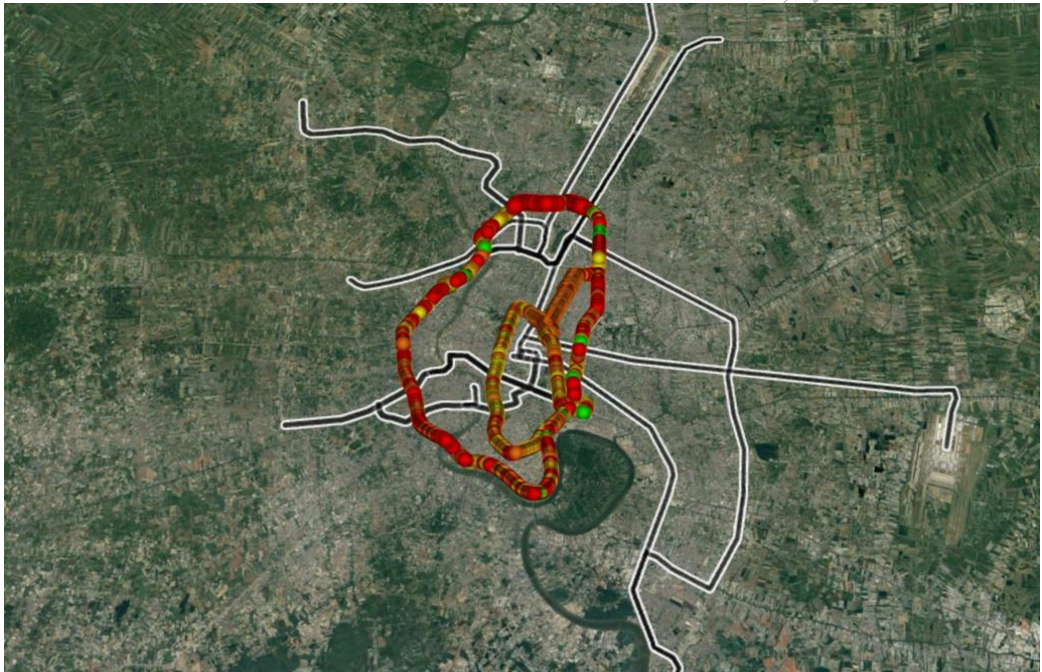

Figure 2. Two routes for the drive test (routes 1 and 2 are in 'the inner ring' and 'outer ring' respectively)

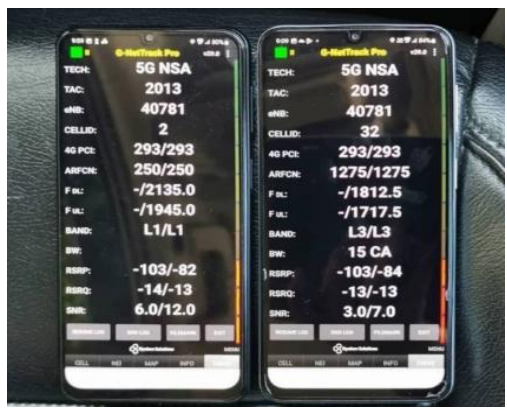

Figure 3. The 5G smartphones utilized in this study, adopted from [8].

## 2.2 Stationary Tests

Stationary tests were conducted as an additional study, focusing on data speeds, DL, and UL. Based on the same 5G smartphones used for the drive tests, the Speedtest application, which is widely used in [23], [29], [30], [34]-[36], was installed in both smartphones. The stationary tests with two MNOs, TRUE and DTAC, were conducted randomly at the first floor (indoor) within the same building as mentioned in [37] called the Rajavithi Hospital during the working hours on December 27, 2023, and the outdoor within one of the most iconic landmarks as mentioned in [38] called the Temple of the Emerald Buddha located in the Grand Palace during the working hours on December 28, 2023, which are very crowded daily. After gathering the data from the stationary tests, the outliers, including minimum and maximum DL speed values, were discarded. Then the data of about 70 records per MNO from both places were presented in the next section. Furthermore, the statistical analysis was also performed.

## 3. RESULTS AND ANALYSIS

Unlike previous studies in other countries that used G-NetTrack and Speedtest applications for mobile network evaluation and similar purposes [23], [26]–[36], this study conducted mobility and stationary tests and utilized these tools for investigating the status of the technical merger between two old operators. The results and the analysis can be described, respectively.

### 3.1 Mobility Test Results

After conducting the drive tests with an average speed of $54.50\pm14.62$ km/h on R1 (the inner ring) and R2 (the outer ring), the raw data were gathered and then combined. Each dataset from each MNO has more than 10,400 records. Then, they were processed, following the flow as shown in Figure 4. Next, the mean (or average) values and standard deviation of three QoS parameters are presented in Table 2. As presented in the table, the results can be described as follows:

- TRUE shows a slightly higher average RSRP value than DTAC. Where, the DTAC's RSRP value is $-82.2\pm9.8$ dBm, whereas the average TRUE's RSRP value is $-81.9\pm10.5$ dBm. The RSRP values from both MNOs are categorized as 'good'.
- TRUE shows a better average RSRQ value of signal than DTAC. The TRUE's RSRQ value is $-12.2\pm2.6$ dBm, while DTAC's RSRQ value is $-14.3\pm3.7$ dBm. The RSRQ from TRUE is classified as 'fair' but DTAC's RSRQ is 'poor'.
- Like RSRQ, average TRUE's SNR value is better than DTAC's value. The TRUE's SNR value is $6.8\pm6.4$ dBm while the DTAC's value is $3.8\pm6.3$ dBm. The SNR values from both MNOs are categorized as 'good'.

It is known that 5G technology must be interoperable with 4G or LTE when 5G network mode is not available. Therefore, the connected technologies or network modes were also investigated from the data obtained from the drive tests and then presented in Figure 5. As shown in the Figure 5(a), the smartphone under the drive tests connected to the TRUE-5G network at only 23.12%, while it connected to the LTE network at 76.88%. It is consistent with the DTAC networks that the phone connected to the 5G network only at 22.01%, while it connected to the LTE network at 77.99%, as presented in Figure 5(b).

Furthermore, from the mobility tests on both routes with the average, the data were analyzed and found that TRUE connected to 139 (about 51%) base stations, while DTAC connected to 104 (about 38%) base stations, whereas there were 31 (about 11%) base stations that TRUE and DTAC used together, or called roaming, during the tests. Therefore, this information can be presented as shown in Figure 6.

### 3.2 Stationary Test Results

After conducting indoor-stationary tests within a building and around the great landmark that are very crowded, the average values of DL and UL speeds after the pre-process were presented in Table 3. As shown in the table, the DL of $210.0\pm108.8$ Mbps provided by TRUE are slightly higher than DL of $199.2\pm99.8$ Mbps provide by DTAC. On the other hand, the UL of $41.7\pm11.3$ Mbps provided by DTAC is higher than UL of $38.0\pm14.5$ Mbps provided by TRUE. Lastly, for latency, as shown in Table 3, TRUE shows lower latency than DTAC. TRUE's latency values is $19.1\pm3.7$ ms, while DTAC's latency is $17.0\pm2.4$ ms.

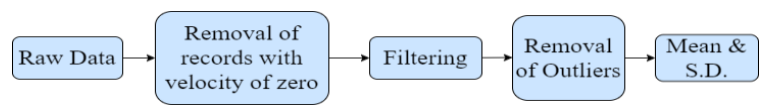

Figure 4. Processes for obtaining the results from the raw data

### 3.3 Analysis

In order to investigate if each pair of QoS parameters is different or not, although some pairs are not very different, a statistical technique called t-Test as used in [21], [39] was applied. Six hypotheses called H1-H6, were also conducted for hypothesis tests. The combined data from R1 and R2 tests in mobility mode were utilized for H1-H3 hypothesis tests, while the data from stationary tests were utilized for H4-H6 hypothesis tests. H1-H6 are described as follows:

H1: The average RSRP provided by DTAC and TRUE is the same or different.
H2: The average RSRQ provided by DTAC and TRUE is the same or different.
H3: The average SNR provided by DTAC and TRUE is the same or different.
H4: The average DL provided by DTAC and TRUE is the same or different.
H5: The average UL provided by DTAC and TRUE is the same or different.
H6: The average latency provided by DTAC and TRUE is the same or different.

Table 2. Results from the mobility tests

| QoS Parameter | MNO | N | Average | S.D. | Remork |
|---|---|---|---|---|---|
| RSRP | DTAC | 10210 | -82.2 | 9.8 | Good |
|  | TRUE | 10043 | -81.9 | 10.5 | Good |
| RSRQ | DTAC | 10210 | -14.3 | 3.7 | Poor |
|  | TRUE | 10043 | -12.2 | 2.6 | Fair |
| SNR | DTAC | 10172 | 3.8 | 6.3 | Good |
|  | TRUE | 10029 | 6.8 | 6.4 | Good |

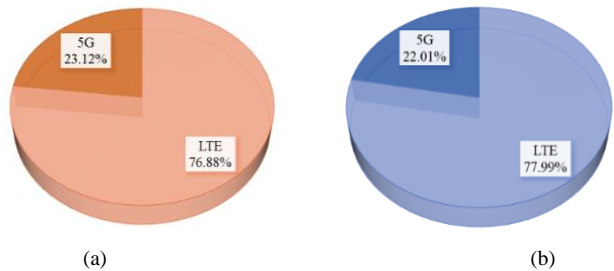

(a)　　　　　　　　　　　　(b)

Figure 5. The portion of 5G and LTE services captured from the drive tests when (a) TRUE and (b) DTAC

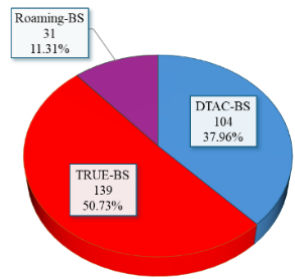

Figure 6. The information of BS stations obtained gathered from the drive tests

Table 3. Results from the stationary tests

| QoS Parameters | MNO | N | Average | S.D. |
|---|---|---|---|---|
| DL | DTAC | 68 | 199.2 Mbps | 99.8 |
|  | TRUE | 70 | 210.0 Mbps | 108.8 |
| UL | DTAC | 73 | 41.7 Mbps | 11.3 |
|  | TRUE | 70 | 38.0 Mbps | 14.5 |
| Latency | DTAC | 71 | 17.0 ms | 2.4 |
|  | TRUE | 66 | 19.1 ms | 3.7 |

Table 4. Results from the analysis using t-Test

| Hypothesis | p-values | Remarks |
|---|---|---|
| H1 | 0.035* | RSRP values from DTAC and TRUE are significant different. |
| H2 | <0.001* | RSRQ values from DTAC and TRUE are significant different. |
| H3 | <0.001* | SNR values from DTAC and TRUE are significant different. |
| H4 | 0.542 | DL values from TRUE and DTAC are insignificant different. |
| H5 | 0.090 | UL values from TRUE and DTAC are insignificant different. |
| H6 | <0.001* | Latency values from DTAC and TRUE are significant different. |

Remark: * means significant at p-value < 0.05 for 95% confidence interval

The results of the hypothesis tests for hypotheses H1 through H6 are displayed in Table 4. The analyzed results show that each pair of RSRP, RSRQ, SNR, and latency provided by DTAC and TRUE is significant, and the p-values are less than 0.05. Whereas the analysis results confirm that average DL speeds gathered from the same place provided by both MNOs are insignificant differences (p-value = 0.542). It is consistent with the hypothesis test results for UL, which are also insignificantly different (p-value = 0.090).

## 4. DISCUSSION AND CONCLUSION

The results from this study show that even though DTAC and TRUE have become the same company already, the mobile networks have not completely merged as presented in the media yet. They operated together by roaming or sharing resources, such as base stations. The empirical evidence from this study shows that overall TRUE is slightly better than DTAC in terms of DL speed, RSRP, RSRQ, and SNR, except UL speed and latency. However, the objective of this study was not to compare the performance between DTAC and TRUE, since both are the same company already. It is to investigate if they have merged in terms of technical already or not. Thus, there are interesting issues that can be discussed.

Firstly, for the drive tests, the routes for the drive tests conducted within the urban areas of Bangkok do not cover the whole area of Bangkok. The tests were conducted between one night and early morning with limited time. Thus, the results obtained from the tests cannot be representative of the results from the whole of Bangkok, although this test mode can cover wider areas than the stationary mode.

Besides, the results (e.g., RSRP, RSRQ, and SNR) from the G-NetTrack application used in this study are too technical for general users. Nevertheless, from the mobility study, as observed and investigated, both smartphones were offered 5G NSA only about 22%–23%, while they connected LTE networks 77%–78%. In fact, DTAC does not support 5G SA services for customers, whereas TRUE announces that it provides 5G SA if devices support it. However, during the drive tests, the smartphone did not get any 5G SA service. It may require making a special request to the TRUE support team to enable the service if it is needed.

For stationary tests, the download and upload speeds, which are user-friendly QoS parameters, from DTAC and TRUE are not significantly different after being investigated by using t-Test. However, the speed rates are very low when compared to the performance of 5G, as it is mentioned that its peak speed is 20 Gbps maximum. Moreover, when considering the average DTAC's latency of about 17 ms and the TRUE's latency of about 19 ms, they are very far from the 1 ms that is the latency of 5G as defined theoretically. This issue may be due to several reasons, such as the default setting mostly used in the Speedtest application and the limitations of the 5G NSA, which still uses some parts of LTE for providing 5G to users. Furthermore, the business management and marketing reasons may be the major reasons, since providing full 5G or high performance of 5G to users requires a high cost of investment from each MNO.

Moreover, according to the objectives of this study for investigating the situation of the technical merger between the old MNOs, the findings from this study can answer the issue. The old customers from DTAC and TURE are provided with 5G mobile phone signals and service separately, except in the case of roaming. In addition, the results and the analyzed results, as shown in Tables 3 and 4, are evidence that the old TRUE customers tend to have a better level of quality of service than the old DTAC customers.

Lastly, for the major findings of this study, not only the disclosure of the status of the technical merger between two old MNOs—which means for the customers of those old operators—but also the methodologies as presented in this paper can also be an example that utilizes the G-NetTrack Pro application as a tool for drive tests while applying the Speedtest application as a standard tool for stationary tests. Both tools can capture a lot of interesting QoS parameters, including RSRP, RSRQ, SNR, DL, UL, and latency.

## ACKNOWLEDGEMENTS

Grateful to Southeast Asia University, Rajamangala University of Technology Phra Nakhon, and Phetchabun Rajabhat University and South - East Asia University for supporting this study. Lastly, thanks to Mr. Peter Bint for English editing.

## BIOGRAPHIES OF AUTHORS

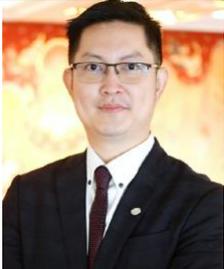
**Surachai Chatchalermpun** is a special lecturer in Computer Engineering, Faculty of Engineering, South - East Asia University. He received first class honors in a B.S. degree in computer engineering, King Mongkut's University of Technology Thonburi (KMUTT), Bangkok, Thailand in 2008. He is an expert on Cybersecurity & Data Privacy and Risk Management, he is now a Country Cyber Security & Privacy Officer (CSPO) at Huawei Technologies (Thailand) Co., Ltd. He was a CISO at Krungthai Bank, and a Regional Head of IT security at Maybank in Asia-Pacific. Also, he holds many international certificates (e.g., CISSP, CISA, CISM, ISO27001 and MIT and Harvard executive certificates). His research interests include cybersecurity, 5G cyber attacks, cloud security, data privacy and data protection. He can be contacted at email: surachai.won@gmail.com.

**Therdpong Daengsi** is an Assistant Professor in the Faculty of Engineering, RMUTP. He received a B.Eng. in Electrical Engineering from KMUTNB in 1997. He received an M.Sc. in Information and Communication Technology from Assumption University in 2008 before receiving a Ph.D. in Information Technology from KMUTNB in 2012. He also obtained certificates including Avaya Certified Expert – IP Telephony and ISO27001. With 19 years of experience in the telecom business sector, he also worked as an independent academic for a short period before becoming a full-time lecturer. His research interests include VoIP, QoS/QoE, mobile networks, multimedia communications, cybersecurity, data science, and AI. He can be contacted at email: therdpong.d@rmutp.ac.th. (Noted: he is also the co-first author and co-corresponding author for this paper.)

**Pakkasit Sriamorntrakul** is now a master's student in the Faculty of Engineering, RMUTP. He received a B.Eng. in Computer Engineering from Mahidol University in 2005. He obtained the Avaya Certified Expert Certificate and was the Avaya Certified Support Specialist in IP Telephony. He also holds other certificates, including Cisco Certified Network Professional, Microsoft Certified Systems Administrator, and VMware Certified Professional 5. He has 18 years of experience in system, network, and telecom businesses. His research interests include high-performance computer systems and networks, VoIP quality measurement, security, mobile network, AI, and IoT. He can be contacted at email: pakkasit-s@rmutp.ac.th.

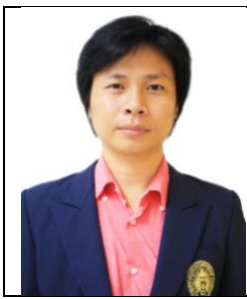 **Kritphon Phanrattanachai** Kritphon Phanrattanachai is an Assistant Professor in the Faculty of Agricultural & Industrial Technology, Phetchabun Rajabhat University (PCRU), Thailand. He received the BSc. degree in electrical industrial from Phetchabun Rajabhat University, Thailand, in 2002. He received an MSc degree in electrical technology from KMUTNB in 2009 and a Ph.D. in Tech.Ed. from KMUTNB in November 2019. Also, he is now an assistant president of PCRU. His research interests include circuit synthesis, simulation of linear and non-linear circuits and systems, IoT, QoS/QoE, mobile networks, telecommunications. He can be contacted at email: kritphon.ai@pcru.ac.th.